 \documentclass[manuscript]{aastex61}
\usepackage{graphicx}
\submitjournal{ApJ}

\shorttitle{\ A multiwavelength study of the Crab Nebula}
\shortauthors{Dubner et al.}

\def\d{$^\circ$}
\def\m{$^\prime$}
\def\s{$^{\prime\prime}$}
\def\hh{$^{\mathrm h}$}
\def\mm{$^{\mathrm m}$}
\def\ss{$^{\mathrm s}$}
\def\cm3{cm$^{-3}$}

\def\2{$^{12}$CO}
\def\3{$^{13}$CO}

\begin{document}

\title{Morphological properties of the Crab Nebula: a detailed multiwavelength study based on new VLA, {\it HST}, {\it Chandra} and {\it XMM-Newton} images.}

\correspondingauthor{Gloria Dubner}
\email{gdubner@iafe.uba.ar}

\author{G. Dubner}
\affiliation{Universidad de Buenos Aires. Facultad de Cs. Exactas y Naturales. Buenos Aires, Argentina}
\affiliation{CONICET-Universidad de Buenos Aires. Instituto de Astronom\'{\i}a y F\'{\i}sica del Espacio (IAFE) \\
CC 67, Suc. 28, 1428 Buenos Aires, Argentina}

\author{G. Castelletti}
\affiliation{Universidad de Buenos Aires. Facultad de Cs. Exactas y Naturales. Buenos Aires, Argentina}
\affiliation{CONICET-Universidad de Buenos Aires. Instituto de Astronom\'{\i}a y F\'{\i}sica del Espacio (IAFE) \\
CC 67, Suc. 28, 1428 Buenos Aires, Argentina}
\nocollaboration

\author{O. Kargaltsev}
\affiliation{Department of Physics, George Washington University \\
 Washington DC 20052, USA}
\nocollaboration

\author{G. G. Pavlov}
\affiliation{Department of Astronomy and Astrophysics \\
Pennsylvania State University, University Park, PA 16802, USA}
\nocollaboration

\author{M. Bietenholz}
\affiliation{Hartebeesthoek Radio Astronomy Observatory \\
P.O. Box 443, Krugersdorp 1740, South Africa}
\affiliation{Department of Physics and Astronomy, York University \\
Toronto, M3J 1P3 Ontario, Canada}
\nocollaboration

\author{A. Talavera}
\affiliation{XMM-Newton Science Operations Centre, ESA \\
Villafranca del Castillo, Apartado 78, E-28691, Villanueva de la Ca\~nada, Spain}

\begin{abstract}

We present a detailed analysis of the morphological properties of the Crab Nebula across the electromagnetic spectrum  based on new and previous high-quality data from radio to X-rays. In the radio range we obtained an image of the entire nebula at 3~GHz with subarcsecond angular resolution using  the VLA (NRAO)  and an image  at 100~GHz of the central region using the ALMA  array. Simultaneously with the VLA observations we  performed  {\it HST} WFPC3  near infrared ($\lambda \sim 1.5 \mu$m) and  {\it Chandra} X-ray (0.5--8 keV band) observations of the central  region of the nebula. In addition we produced a new UV image of the Crab nebula at 291 nm by co-adding 75 individual exposures of the Optical-UV Monitor on board  \it XMM-Newton\rm. The high-angular resolution and 
high-dynamic range  radio image at 3 GHz allowed us to improve the detection and characterization of  peculiar morphological features including arches with foot brightening  and intercrossed loop-like structures, likely  originating in plasma confined to magnetic field lines.
   Based on the new radio image, we carried out a detailed multiwavelength correlation. In the central area, the comparison of  the almost simultaneous  images  confirms that the wisps in the three spectral ranges  do not generally coincide in location,  the radio emission being the most discordant, which is suggestive of  the existence of two different synchrotron components. The X-ray pulsar jet  does not have a  radio counterpart. Instead, another jet-like feature is seen in radio, though with different curvature and starting point.

\end{abstract}

\section{Introduction} \label{introduction}

The Crab Nebula is the archetype of the nebula created through the interaction of the ultrarelativistic wind  injected by a rapidly rotating neutron star, the so-called pulsar wind nebula (PWN). As a very bright object located at a  distance of 2~kpc \citep{trimble1973}, it is an advantageous target to investigate physical processes in PWNe as well as in supernova remnants (SNRs). Thousands of articles have been published reporting multiwavelength observations and theoretical models describing and interpreting the Crab Nebula. Yet, in spite of being one of the best studied objects beyond our own solar system, there is still much to learn about it.

The wealth of existing data from low radio frequencies to very high-energy gamma rays reveals a very rich and dynamic structure  created by the pulsar wind expanding through the remainings of a stellar explosion. The  reviews  by \citet{hester2008} and \citet{buhler2014} provide extensive overviews of  the current knowledge on  this source.
In a simplified picture, there are three different components from inside out: (i) the 33~millisecond pulsar  PSR~B0531+21 (J0534+2200)  powering the nebula, (ii) the synchrotron-emitting shocked pulsar's wind, highly dynamic and with a wealth of fine-scale structure, and  (iii) a network of thermal filaments from the ejecta (and possibly some material from the precursor stellar wind) compressed by the PWN. A fourth component would be  a much larger, almost unseen, freely expanding SNR, claimed to have been  detected in the UV by means of C~IV $\lambda$1550 absorption \citep{sollerman2000}. However, the  outer shock of the expanding SNR has not yet been detected  despite considerable searching efforts.   In what follows we summarize the main observational characteristics of the Crab Nebula in  different spectral ranges between radio and X-rays.

The synchrotron nebula expands within the ``cage''  of filaments beautifully displayed in emission lines in visible light
\citep[e.g.][]{hester1995,hester1996,sankrit1997,loll2013}. Most  of the emission from the  filaments is the result of photoionization  by the hard continuum from the synchrotron PWN emission; there is also emission arising from radiative cooling behind the shock driven by the PWN into the freely expanding ejecta \citep{sankrit1998}. The edge of the synchrotron nebula moving outward faster than the local expansion velocity has swept up what is called the ``skin,'' a thin shell of thermal material that apparently marks the local boundary of the synchrotron nebula.  The center of the nebula is characterized by a torus  that surrounds the pulsar in the equatorial plane of the neutron star and  a jet aligned with the spin axis of the pulsar. These features  are highly dynamic. The most prominent features in the torus move outward with speeds of up to $\sim 0.5c$ as  observed in optical, infrared, and in X-rays \citep[][and references therein]{hester2008}. Some synchrotron features in the torus, called wisps,\footnote{By convention the optical features seen in the synchrotron nebula are called ``wisps" (other identified features are ``sprite" and ``knots"), while the thermal gas structures seen in emission lines are called ``filaments"  \citep{hester2008}.}  are also seen in radio, but, as described below,  they do not match up well with the wisps seen in visible light and in X-rays.

In the infrared (IR) domain several observations of the Crab Nebula have been conducted both from Earth in the near-IR \citep[e.g.][]{sollerman2003, tziamtzis2009} and from space in mid- and far- IR  using {\it IRAS} \citep {marsden1984}, {\it ISO} \citep{douvion2001, green2004}, {\it Spitzer} \citep{temim2006, temim2012}, and {\it Herschel} \citep{gomez2012}.  \citet{persi2012} presents a review of previous IR studies. The emission of the Crab Nebula in the near- and mid-IR is dominated by continuum synchrotron radiation, while the  emission observed in the far-IR (24 and 70~$\mu$m {\it Spitzer} images)  may be due to line emission or to the presence of a small amount of warm dust \citep{temim2006}. In the FIR domain, the Crab Nebula has  also been explored using PACS (wavelengths from 70 to 160~$\mu$m) and SPIRE (from 250 to 500~$\mu$m) on board of {\it Herschel}. Based on these observations,  \citet{gomez2012} reported the existence of a cool dust component along the ionized filamentary structure, spatially coincident with the location of ejecta material.

In the X-ray range, the Crab Nebula was one of the first discovered extrasolar  sources, when  in the early stages of the X-ray astronomy it was detected with a rocket-borne experiment \citep{gursky1963, bowyer1964}. Over the years the pulsar and the nebula have been  observed using almost all available instruments  \citep[see e.g.][and references therein]{weisskopf2000, mori2004, seward2006, madsen2015}. The  X-ray remnant is  smaller than the optical and IR nebula, which is reasonably explained by the fact that the X-ray emitting electrons have a shorter lifetime than the lower energy electrons emitting optical and IR radiation.  In the 1-10~keV band only  $\sim$5\% of the radiation comes from the pulsar, while most of the nebular X-ray emission originates in  the equatorial torus and jet
\citep{brinkmann1985, hester1995}. \citet{seward2006} reported {\it Chandra} observations revealing for the first time the existence of faint structures forming fingers, loops, and bays in all directions beyond the bright torus-jet structure, whose spectra soften with the distance from the pulsar. The authors suggest that the structure is determined by the synchrotron lifetime of diffusing particles.

In the radio band, there are studies of the Crab Nebula since the early 1950s
\citep [see e.g.][etc.]{mills1952, baars1972, duin1972, wilson1972, weiler1975, swinbank1979, velusamy1984, velusamy1992} reporting total intensity and polarization  measurements
made at several frequencies  with different instruments and various angular resolutions. Those early studies revealed an ellipsoidal source, somewhat larger than the optical remnant, with a filamentary appearance and a complex polarization structure.
\citet{bietenholz1991} conducted VLA (NRAO) radio observations at 1.4 and 5~GHz, presenting polarization and Faraday rotation maps with the angular resolution of 1\farcs8. Also, \citet{bietenholz1991b} used two-epoch high-resolution radio data  to investigate the expansion of the outer edge of the nebula, finding that the expansion has accelerated since the supernova explosion.  The center of the nebula and its dynamical changes were investigated in great detail in  radio frequencies \citep{bietenholz1992, bietenholz2001, bietenholz2004}, demonstrating that the radio wisps form elliptical ripples  similar to the optical wisps in both morphology and behavior. However, from the comparison of radio to simultaneous optical images of the central part of the Crab Nebula, it is  found   that the radio wisps are sometimes displaced from the optical ones or have no optical counterparts, and the brightest optical wisps near the pulsar do not seem to have radio counterparts.  On the contrary, in the exterior of the nebula  there is a good general correspondence  between the radio and optical features  \citep{bietenholz2004}.

In addition to the mentioned morphological characteristics, there can also be noted the  ``synchrotron bays,''  large indentations on the periphery of the synchrotron  nebula (one very clear to the east and the other somewhat less conspicuous to the west), a peculiarity that is evident in the optical continuum, UV, and  IR images, but is less clear in radio and in  X-ray   \citep{wilson1972, seward1989}. High linear polarization (between 30\% and 60\%) with radial orientation  has been observed along the edges of the bays. The absence of synchrotron emission within the bays has been interpreted by invoking a magnetized torus encircling the remnant (probably formed by  a disklike ejection from the progenitor star) that blocks the penetration of the relativistic particles from the pulsar  creating these dark regions \citep {fesen1992}.

As a part of a larger program dedicated to the investigation of the broad spectrum of the wisps and other central features near the pulsar,  aiming at tracing the  spectral energy distribution of the injected particles \citep{krassilchtchikov2014}, we have undertaken a new detailed, almost contemporaneous, study of the Crab Nebula using the NRAO Karl Jansky Very Large Array (VLA)\footnote{The Karl G. Jansky Very Large (EVLA) is
operated by the National Radio Astronomy Observatory (NRAO), a facility of the National Science Foundation operated under cooperative agreement by Associated Universities, Inc.}  at 3~GHz, the \it Hubble Space Telescope (HST) \rm and {\it Chandra} telescopes. We have also used the Atacama Large Millimeter/submillimeter Array (ALMA) to produce the first detailed radio continuum image of the center of the nebula  at 100~GHz, although the ALMA observations were not contemporaneous with the others. In this work we present the new 3~GHz radio image  and first results of the ALMA
100~GHz observations,  a new high spatial resolution UV image of the entire Crab Nebula obtained from data acquired by the Optical Monitor of XMM-{\it Newton} ($\lambda \sim$ 291 nm), a new {\it HST} NIR image ($\lambda \sim 1.5 \mu$m)  and a new {\it Chandra} X-ray image in the 0.5--8.0 keV band. This paper is devoted to the detailed analysis of the morphological properties of  the Crab Nebula across the electromagnetic spectrum. To carry out this comparison, in addition to our own new data we used existing {\it Spitzer} and {\it HST} images of the entire remnant kindly provided by T. Temim \citep[data from][]{temim2006} and by A. Loll \citep[data from][]{loll2013}, respectively, which in spite of not being contemporaneous,  are  nevertheless useful  for a comparative analysis of the  larger scale structure.

\section{Observations and data reduction}
\label{observations}

\subsection{VLA radio data}
Radio observations of the Crab Nebula were performed with the VLA of the NRAO
on 2012 November 26 and 27 (Observing code VLA/12B-380, Legacy project AD670).
The data set was acquired in the  A-array configuration of the interferometer using the broadband VLA S-band receiver.
The observations used two 1024~MHz independent basebands (covering a total bandwidth between 1988 and 3884~MHz) each with eight 128~MHz contiguous subbands,
spread into 64 channels each.
 On both days we observed the source J0137+3309 (3C~48) as flux density and bandpass calibrator,
setting the flux scale according to the coefficients derived at the EVLA by NRAO staff in 2010.
Regular observations on the source J0559+2353 were used for phase calibration.

The raw visibilities were calibrated and imaged using the NRAO Common Astronomy Software Applications  (CASA) package.
Data from each day were first independently calibrated and then combined into a
single \it uv \rm data set to construct the final image.
After initial data editing we determined the antenna gains across the bandwidth for flux and phase calibrators. Due to the presence of strong radio frequency interference, two
spectral windows were dropped  in both observing days.
Before solving for the complex bandpass shape, we first estimated the relative delays
of each antenna relative to the reference antenna.
In the next step the calibration solutions were applied to the target source, which
was then split out from the data set. We made the image using multi-frequency
synthesis (MFS) and multi-scale (MS)  CLEAN as implemented in CASA.
This procedure accounts
for variations in the synthesized beam while it solves for the scale size of the
emission across the field of view. We set the  Briggs robust parameter
equal to 0.5, which represents a compromise between natural and uniform weighting. We then performed two rounds of self-calibration using the CLEAN models.

\begin{figure}
\includegraphics [scale=0.7]{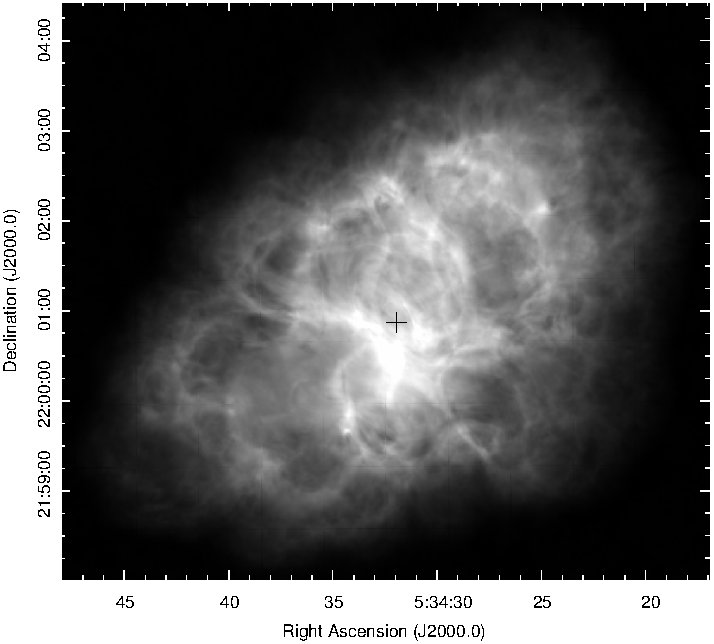}
\caption{VLA radio image of the Crab Nebula at 3~GHz. It has HPBW = 0\farcs93 $\times$ 0\farcs80 at PA=76$^\circ.$3 and a noise level of
0.03~mJy~beam$^{-1}$. It is displayed in a linear scale between 0.09 and 17
mJy~beam$^{-1}$. The plus sign shows the location of  PSR~B0531+21 (J0534+2200).
\label{radio-complete}}
\end{figure}

Observations in the A-array around 3~GHz provide information on spatial scales between the theoretical synthesized beam of $\sim$0\farcs7 and the largest angular scale of $\sim$18\s~ that can be well imaged  at this frequency. This range allows us to map in high detail the thin radio wisps and synchrotron knots, but misses the extended features.  To produce an image with the large-scale contributions, we followed a strategy somewhat similar to that described in  \citet{bietenholz2004} using the same 5 GHz image a template after correcting it for spectral differences in brightness to a frequency of 3 GHz and taking into account a long-term decay at a mean rate of 0.20 \%~yr$^{-1}$ \citep{aller1985, vinyaikin2007}. The resulting model was also spatially scaled to account for the nebular expansion at a rate of 0.13 \%~yr$^{-1}$ \citep{bietenholz2015}. The CASA task FEATHER was used to combine the different spatial resolution images, using their Fourier transforms. The  final 3 GHz image recovers a total flux density of  743 Jy and has a  synthesized beam  0\farcs93$\times$0\farcs80 at PA=76$^\circ$.3 and an rms noise level= 0.03~mJy~beam$^{-1}$. The image produced with the contribution of all spatial scales recovered is presented in Figure~\ref{radio-complete}.

Since the driving purpose of this work is to perform a detailed, comparative morphological analysis of the Crab Nebula across the electromagnetic spectrum,  based on the same VLA data we produced another 3 GHz radio image using a lower weight factor for the contribution of the scaled lower resolution data. In essence this image is high-pass filtered, and some of the structure  at larger spatial scales has been removed. At  the cost of losing some smooth extended emission and not recovering the total flux density, the contrast is  improved and all the thin structure, including narrow filaments and wisps, are better revealed.  In the rest of the paper, we use this last image (shown in Figure~\ref{radio}) to match and analyze the 
small-scale structure.

\subsection{ALMA radio data}
To survey the central portion of the Crab Nebula at millimeter wavelengths, we used the 
ALMA 12 m array in the C32-4 configuration for a total of 2.2~hr and  the ALMA compact array (ACA, A7 m) for a total of 6.6~hr in  Band 3 (around 100~GHz, 3.3~mm) (Cycle 1, Program 2012.1.01099.S). The observations took place in blocks during several sessions between 2013-10-06 and 2014-05-04 (Cycles 1 and 2). The region observed was a rectangle  1\farcm5$\times$1\farcm1 in size centered at
05$^{\mathrm{h}}$34$^{\mathrm{m}}$31$^{\mathrm{s}}$,
+22$^{\circ}$00$^{\prime}$52$^{\prime\prime}$ and tilted about 70\d~  east of north. The area was surveyed through a mosaic of seven individual pointings separated by the Nyquist criterion. For the two arrays, 30 and 8 antennas were, respectively used. The sources J0423-1200, J0510+1800 and J0521+1638 were used for calibration. The correlator was configured to observe four 2000 MHz wide windows  centered at 93, 95, 105, and 107~GHz, each divided in 128 channels spaced in 15.625~MHz.

Automated pipeline processing was used to produce the images. The final image obtained with ACA 7 m array has HPBW = 11\farcs9 $\times$ 11\farcs6, and for the 12 m array  HPBW = 1\farcs8 $\times$ 1\farcs4 and rms noise of $\sim$ 0.2~mJy~beam$^{-1}$.  The ACA observations were carried out to provide the missing short spacings contribution; however, the procedure was not completed  because the 12 m array observations do not meet the quality expectations and the rms achieved was about 10 times worse than requested. The reason for that may be that in an effort to make this a short, fast project, a small mosaic covering only the central $\sim$ 5\% of the Crab Nebula was mapped. This area, although sufficient to match the contemporaneous {\it Chandra} and  {\it HST} studies of the  structures around the pulsar,   turned out to be too small for a very bright radio source such as is the Crab Nebula,  and much extended missing structure was left out, degrading the quality of  the image.  In addition, a problem that can result in  small distortions of the final image has been recently identified affecting  mosaics observed during Cycles 1 and 2. But probably the main factor impeding the production of an optimal quality image is that the ALMA observations were not carried out within a small temporal window in coincidence with the VLA, {\it HST}, and {\it Chandra} campaigns (as originally requested by the authors) but spread over a six-month period. Therefore, in a target with rapidly moving features such as the wisps, which move  with  projected speeds  of 
$\sim$ 0.2c, the features moved about 3\s~($\sim$ two beams) over the time between consecutive observations, thus corrupting the image.

In spite of not achieving the expected quality, we  report here the  image obtained with the 
12 m antennae array, where the central part of the Crab Nebula is shown for the first time at this high radio  frequency with an arcseccond angular resolution.  The main features are easily identified and can be used for visual  morphological comparison.
Because of the lack of short spatial frequencies in this image, there are negative values around the bright features. The new image presented here improves by a factor of 10 the angular resolution obtained at this frequency for the Crab Nebula up to the present \citep{arendt2011}.
The present observations will serve as a pathfinder for the optimally designed, deeper observations of the entire Crab Nebula with ALMA.

\subsection{{\it HST} Near-IR Data}

To accomplish the proposed morphological comparison across the electromagnetic spectrum, we produced a new high-angular resolution image of the Crab Nebula using the {\it HST}. Within a single {\it HST} orbit on 2012 November 26 (program ID 13043) we acquired four  exposures with the WFC3/IR F160W filter (band: 1.4$-$1.7 $\mu$m) at four dither points using the WFC3-IR-DITHER-LINE pattern with the $0\farcs636$ spacing. This sequence uses evenly spaced 50 s time interval between nine reads. The images from the individual exposures  were ``drizzled'' into a single image by the standard WFC3 pipeline processing.   The total accumulated scientific exposure in the combined 123\s $\times$ 136\s~ image used in this paper is  1612 s. The PSF FWHM is 0\farcs15.

\subsection{{\it XMM-Newton} UV data}

In the UV domain we produced a new image  based on existing  data using the broadband UVW1 filter centered at 291 nanometers (254 --328 nm at half maximum) of the Optical Monitor (OM) on board of {\it XMM-Newton}. The OM is a 30 cm diameter telescope that provides simultaneous observations in a 17\m $\times$ 17\m~ field of view. The data were processed using the {\it XMM} Science Analysis Software (SAS).

 Since the Crab Nebula is one of the main calibration sources for the {\it XMM-Newton} EPIC instrument, there is a good collection of images acquired with OM during the whole period that {\it XMM-Newton} has been operational.  {\it XMM} SAS processing corrects every single image for the geometric distortion introduced by the OM detector. This allows us to align and co-add 75 images  spanning 14 yr, between 2001 and 2015 with high accuracy, as it can be seen by looking at the stars in the field of view, which have an FWHM of 1\farcs7 in the co-added image.

\subsection{{\it Chandra} X-Ray Data}

The Crab Nebula was observed with {\it Chandra} on 2012 November 26  (Obs. ID 14458). Due to the X-ray brightness of the Crab PWN, a single-chip graded\footnote{In the graded mode only the total reconstructed energy of an X-ray event
is telemetered to the ground.}  ACIS-S observation (start time 56257.8736 MJD) was conducted with the shortest
possible 0.2 s frame time for the $300\times300$ pixel spatial window.
The dither amplitude was set to 1\s~  to reduce the blurring.
Due to the small frame time and telemetry saturation, the
dead time was about 90\% of the frame time, and the live time of the 10 ks exposure amounted to only 1244 s of useful scientific exposure. The resolution of the resulting 0.5-8 keV images is about $0\farcs4-0\farcs5$ (PSF FWHM), depending on the off-axis distance.

\section {The 3~GHz and 100~GHz Radio Images}
\label{radio-section}

\begin{figure}[ht!]
\plotone{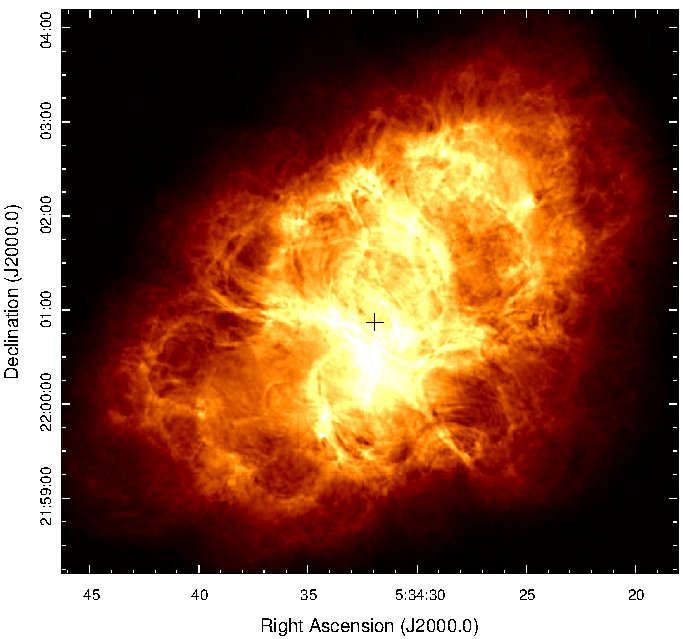}
\caption{ VLA radio image of the Crab Nebula at 3~GHz produced to improve the contrast and better reveal details of small-scale emission as described in Section~\ref{observations}. The beam is 0\farcs93 $\times$ 0\farcs80 at PA=76\fdg3 and the rms noise level is 0.03~mJy~beam$^{-1}$. It is displayed in a linear scale between 0.09 and 5.3 mJy~beam$^{-1}$. The plus sign shows the location of  the pulsar.
\label{radio}}
\end{figure}

In Figure~\ref{radio} we show the high-pass filtered (see Section~\ref{observations}) 3 GHz image of the Crab Nebula, which reveals  the rich structure of the radio synchrotron emission.  In addition to the well-known long, bright filaments, many short structures of different thickness and brightness are evident across the nebula.   The resolved  structure consists of  a tangled net of intercrossed filaments that probably are at different depths along the line of sight. Along the eastern and southern sides, the outer border  of the nebula is clearly undulated, with large arches, as already noticed in the previous radio images of the Crab Nebula.

\begin{figure}
\plotone{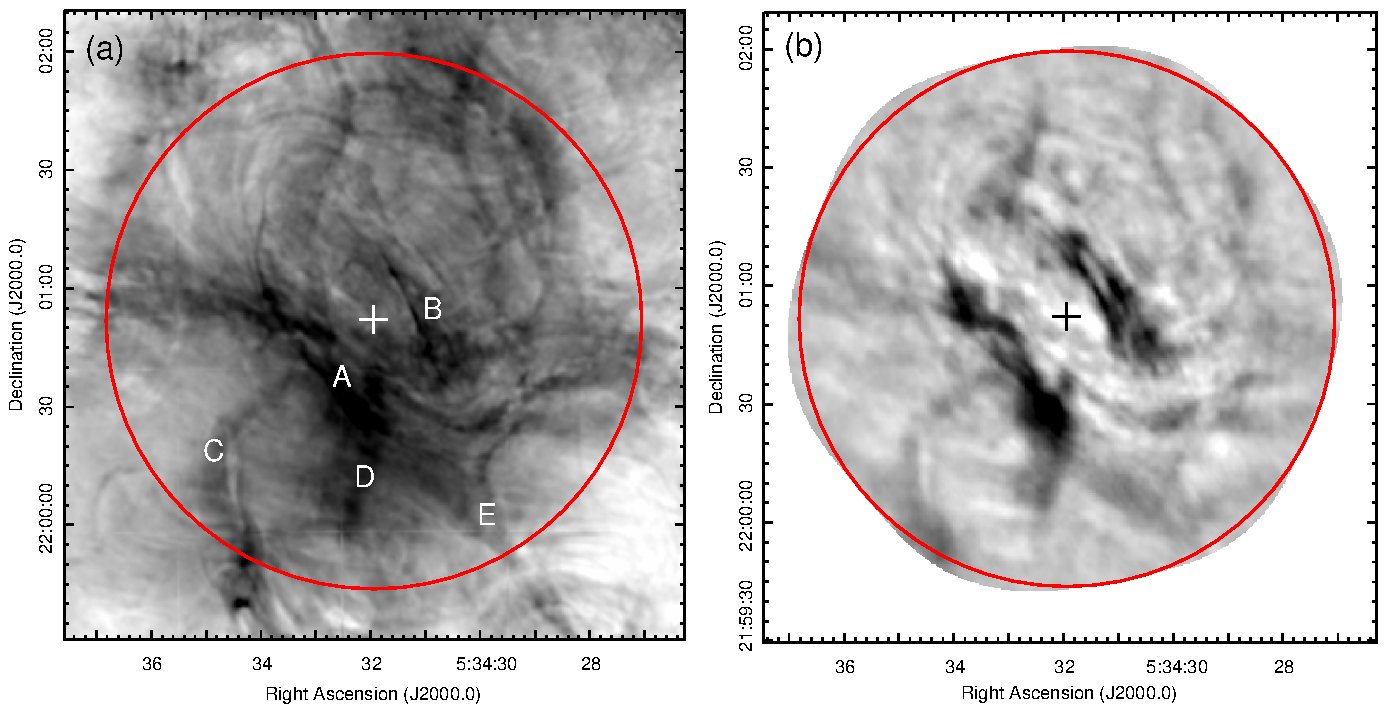}
\caption{{\bf (a) } VLA radio image of the central  $\sim$ 2\m $\times$ 2\m
region of the Crab Nebula  at 3~GHz. The gray scale varies between 1.4 and 5.8~mJy~beam$^{-1}$. The image used is the spatially high-pass filtered as described in Section~\ref{observations} to emphasize the fine structure. The features labeled A, B, C, D, and E are discussed in the text.
{\bf (b)} ALMA radio image at 100~GHz of the same area with HPBW= 1\s.8 $\times$ 1\s.4. The gray scale varies between $-$1.2 and 3.7~mJy~beam$^{-1}$. Negative values are due to the lack of short spacing contributions in the image. A color-inverted display is used to emphasize the prominent features and the plus sign indicates the location of  the pulsar in both panels.
\label{equatorial-region}}
\end{figure}

To analyze the structures in detail,  in Figure~\ref{equatorial-region} we show a close-up of  the equatorial region of the Crab Nebula as seen at 3~GHz and at 100~GHz.   The emission in this region consists   of bright short filaments with different widths and orientations. Two long, almost parallel, bright filaments  (more evident in the image at 100~GHz) are noticed symmetrically located at each side of the pulsar (labeled as features A and B in Figure~\ref{equatorial-region}), at about 5\farcs5  ($\sim 1.6 \times 10^{17}$~cm at 2~kpc) from the pulsar. Another conspicuous feature is the one named C in Figure~\ref{equatorial-region}, with a jet-like appearance, that will be discussed later. From the present radio observations, we can conclude that in the region around the pulsar the spatial distribution at 100 GHz is very similar in appearance to that observed at 3 GHz, although a firm conclusion will only be achieved based on an image at 100 GHz with all spatial scales recovered.

\begin{figure}
\includegraphics[scale=0.5]{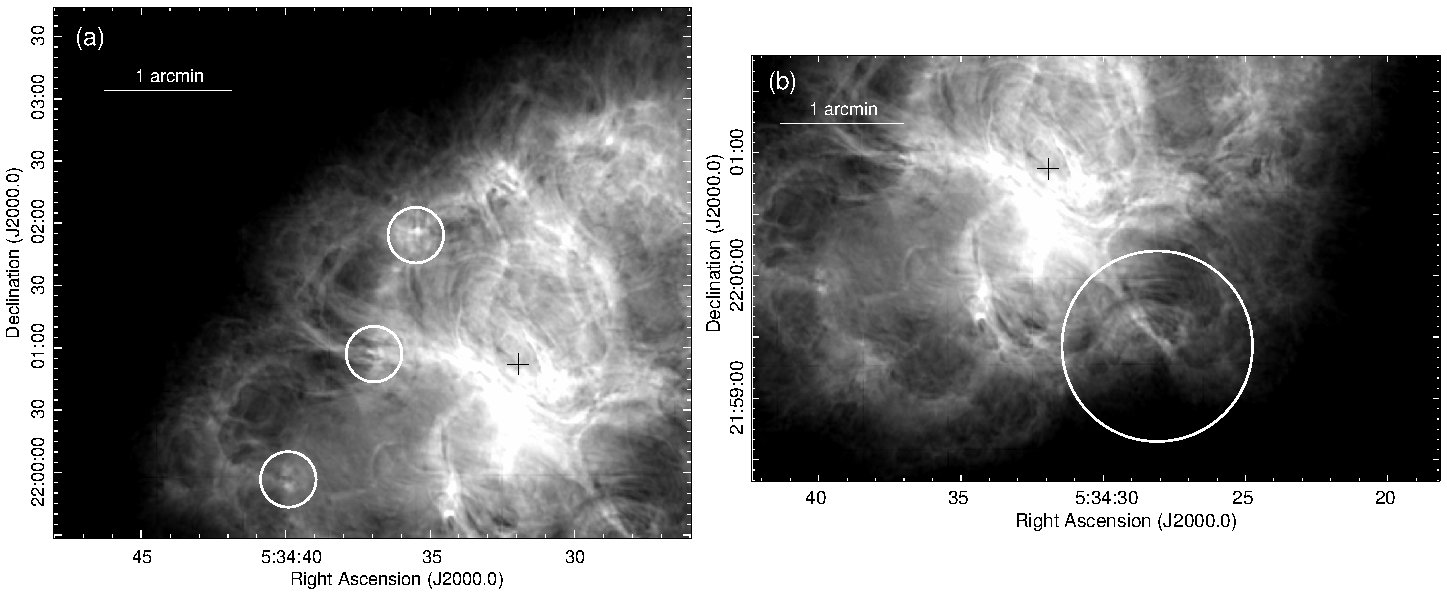}
\caption{{\bf (a)} Close-up view of the eastern flank of the Crab Nebula emphasizing the presence of arches and brightenings at the loop footpoints.  The location of the footpoints are indicated  by small white circles. {\bf (b)} Close-up of the southwestern side. The circle marks the region where an arcade of loops is seen surrounding the filament labeled E in Figure~\ref{equatorial-region}a. For both figures the gray scale is linear and varies between 5 and 50~mJy~beam$^{-1}$.
The plus sign shows the location of the pulsar. The image used is the spatially high-pass filtered one to emphasize the fine structure (see  Section~\ref{observations}).
\label{radio-details}}
\end{figure}

Other interesting features revealed by the 3~GHz observations are seen in
Figure~\ref{radio-details}a, where we displayed a close-up of the eastern side of the nebula. It shows  several peripheral arches \citep[the loop-like features noticed by][]{bietenholz1991} with small bright spots at their footpoints\footnote{ The terminology to name these features is taken from the solar case because of their similar appearance. }
(to guide the eye, we encircled the regions around the footpoints in Figure~\ref{radio-details}a).  The arcs are composed of a series of narrow filaments, well resolved with the present observations, with widths  of the order of 2\s $-$ 3\s~ ($\sim$ 6 to 8 $\times 10^{17}$~cm at 2~kpc),  that rise about 40\s~  above the nebular region, while the separation between the bright spots at the base is of  $\sim$  66\s. Such topology resembles the complex system of arches  observed in solar bursts, where the loops carry plasma confined to magnetic field lines, and the brightening at the footpoints results from particle acceleration and magnetic enhancement
\citep[see for example][]{cristiani2008}. In the  case of solar bursts it is found that the arches reconnect at low heights, and the reconnected arches are anchored in regions that have the magnetic field intensity about an order of magnitude higher than the field in the surroundings. The solar synchrotron radio emission  comes from electrons spiraling down along the reconnected loops. Although the phenomena that give rise to the observed radio  loops in the Crab Nebula have different scales than those acting at the solar bursts, the similarity in the morphology suggests that analogous magnetic loops may be present in the Crab PWN, with the caveat that  the physical parameters (magnetic field strength, plasma density, etc.)  of the respective environments are different by orders of magnitude.  To have an idea, in the solar flares the magnetic fields at the footpoints are of the order of hundreds of Gauss and at the top around 10$-$50 Gauss \citep{tandberg-hanssen2009}, while in the Crab Nebula the magnetic field strength is of about 500 $\mu$Gauss \citep{hester1996}. Also the plasma density in solar flares is typically of the order of $10^9$ to $10^{10}$ cm$^{-3}$ \citep{tandberg-hanssen2009}, while  in PWNe  the ambient number density is of $\sim$ 0.1 cm$^{-3}$ \citep{gaensler2006, reynolds2012}.

In Figure~\ref{radio-details}b we focus on another peculiar feature, which is the set of loops with an arcade-like appearance\footnote{Again, we adopt solar terminology because of the strong resemblance in appearance. Solar arcades are a series of closely occurring loops of magnetic lines of force.} (location marked by a white circle). This structure appears to be formed by  closely packed arches  wrapping around an almost straight bright filament  (named E in Figure~\ref{equatorial-region}a).  Such a topology can be produced if accelerated particles flowing from the pulsar along the filament carry a current, and the arcades are tracing the magnetic field lines  wrapping around (assuming that the magnetic field lines are frozen into the plasma).

In general, one can  identify several sets of closely packed loops with different orientations across the nebula, except for the northwestern edge of the nebula, which looks more open than the rest. It should be noted that these loop-like features appear as a different component, distinct from the wisps and the extensive network of filaments that also shine in other spectral ranges, and that the Crab Nebula is unique among the PWNe observed in our Galaxy  in displaying a morphology with loops at the periphery with such a clearness  \citep[see other PWNe morphologies for example in][]{gaensler2006}.
 \citet{begelman1998}  suggested that kink instabilities can  disrupt the concentric field structure in the nebula and  drive the system toward a more chaotic structure that  may form loops after reconnection.  This hypothesis would be compatible with the widespread synchrotron loops and cellular structure of the magnetic field traced by the synchrotron emission.

In the following sections we analyze the possible correspondences  between the radio emission and the radiation  in other spectral ranges by comparing, for the first time,  data with almost the same angular resolution in all spectral ranges. Before combining all images were regridded to the same projection, coordinate frame, and geometry using AIPS tools. The co-alignment was done using the WCS tools in SAO DS9.

\section {Radio and infrared} \label{infrared}

 As summarized in Section~\ref{introduction} several studies of the Crab Nebula have been performed in the IR range.  Here we have used the 3.6, 4.5, 5.8, 8.0, 24, and 70 $\mu$m images of the Crab Nebula obtained with the {\it Spitzer Space Telescope} IRAC and MIPS cameras 
\citep{temim2006} to compare with the radio emission distribution. At lower wavelengths (4.5 $\mu$m) the emission observed is mostly of synchrotron origin, while at higher wavelengths (8, 24, 70~$\mu$m), the emission is dominated by filaments of forbidden line emission \citep[from S, Si, Ne, Ar, O, Fe, and Ni,][]{temim2012}.

\begin{figure}
\centering
\includegraphics[scale=0.9]{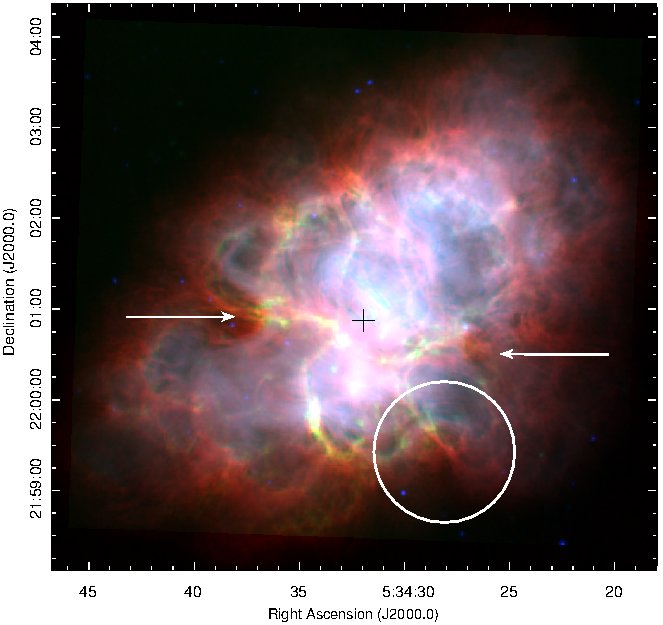}
\caption{Comparison of the radio nebula (in red) with the {\it Spitzer} image at
$\lambda~ 8~\mu$m (in green)  and in 4.5~$\mu$m (in blue).  The white arrows point out the eastern and western ``bays'' discussed in the text. The white circle marks the same region as in Figure~\ref{radio-details}b, where an arcade of loops surrounds filament E. The radio image used  is the spatially high-pass filtered  to emphasize the fine structure (see Section~\ref{observations}).
The plus sign shows the location of the pulsar.
\label{comparison-radio-58-8}}
\end{figure}

In Figure~\ref{comparison-radio-58-8} we present a composite three-color image with radio emission in red, IR at  8~$\mu$m in green, and IR at 4.5~$\mu$m in blue.  In this figure, all the prominent features around the pulsar (labeled from A to D in Figure~\ref{equatorial-region}(a)), are also bright in IR. The interior of the nebula is filled with  synchrotron radiation, seen as bluish  diffuse emission. The surrounding radio emission is traced in red. The IR synchrotron radiation seems to extend further toward the northwest than to the southeast, reaching in the northwest the same extension as the radio emission.

Another  fact that can be seen from this comparison is related to the ``bays''.  As mentioned above, the  east and west bays are two very conspicuous indentations located $\sim$ 80\s~  east (around 5\hh 34\mm 37\ss, +22\d 00\m 57\s) and $\sim$ 75\s~  west of the pulsar  (around 5\hh 34\mm 26\ss, +22\d 00\m 35\s); pointed out by white arrows in Figure~\ref{comparison-radio-58-8}. These cavities were identified at the very discovery of the Crab Nebula in visible wavelengths, and later also detected  in UV and IR. In radio, however, they are not obvious. In Figure~\ref{comparison-radio-58-8}, the superposition of IR and radio  demonstrates that the radio emission follows  the same concave morphology visible in IR, confirming that the bays are also traced in radio wavelengths, where they are somewhat hidden by the bright arched filaments. Particularly, in the case of the east bay, two sets of thin arched filaments  seem to emanate from the center of the bay and curve toward the northwest and toward the southeast.  It can be noticed that the bright footpoints  around  5\hh 34\mm 36\ss, +22\d 01\m 00\s,  mentioned earlier are also bright in IR.  Along the boundary of the bays the radio emission  appears slightly exterior to the IR border, but this can be the consequence of the 10 yr lapse between the observations.

 An additional characteristic from the IR-radio comparison that deserves to be mentioned is related to the  structure that is displayed in Figure~\ref{radio-details}(b) and marked with a white circle in Figure~\ref{comparison-radio-58-8}. It can be noticed that the IR emission perfectly matches in shape and size the radio structure consisting of  closely packed arches wrapping around the straight radio filament  labeled E in Figure~\ref{equatorial-region}(a), which we called the ``arcade'' following the solar nomenclature. The same characteristic will be noticed again in Section\ref{optical}  based on the comparison with the optical continuum emission.

\section {Radio and optical}
\label{optical}

The comparison with the optical emission was carried out in two ways. First, for the complete nebula we used a set of images from the {\it HST} WFPC-2 survey carried out to cover the entire Crab Nebula \citep{loll2013}. The survey includes eight WFPC-2 fields taken between 1999 October and 2002 January in the following filters: F502N ([OIII]), F673N ([SII]),
F631N ([OI]), and F547M (a filter admitting a relatively line-free continuum). Second, to analyze the emission in the pulsar vicinity we used our own contemporaneous {\it HST} images described in Section~\ref{observations}. Although these observation lie in the near-IR range ($\lambda \sim 1.5 \mu$m), we include them here for convenience.

\begin{figure}
\plotone{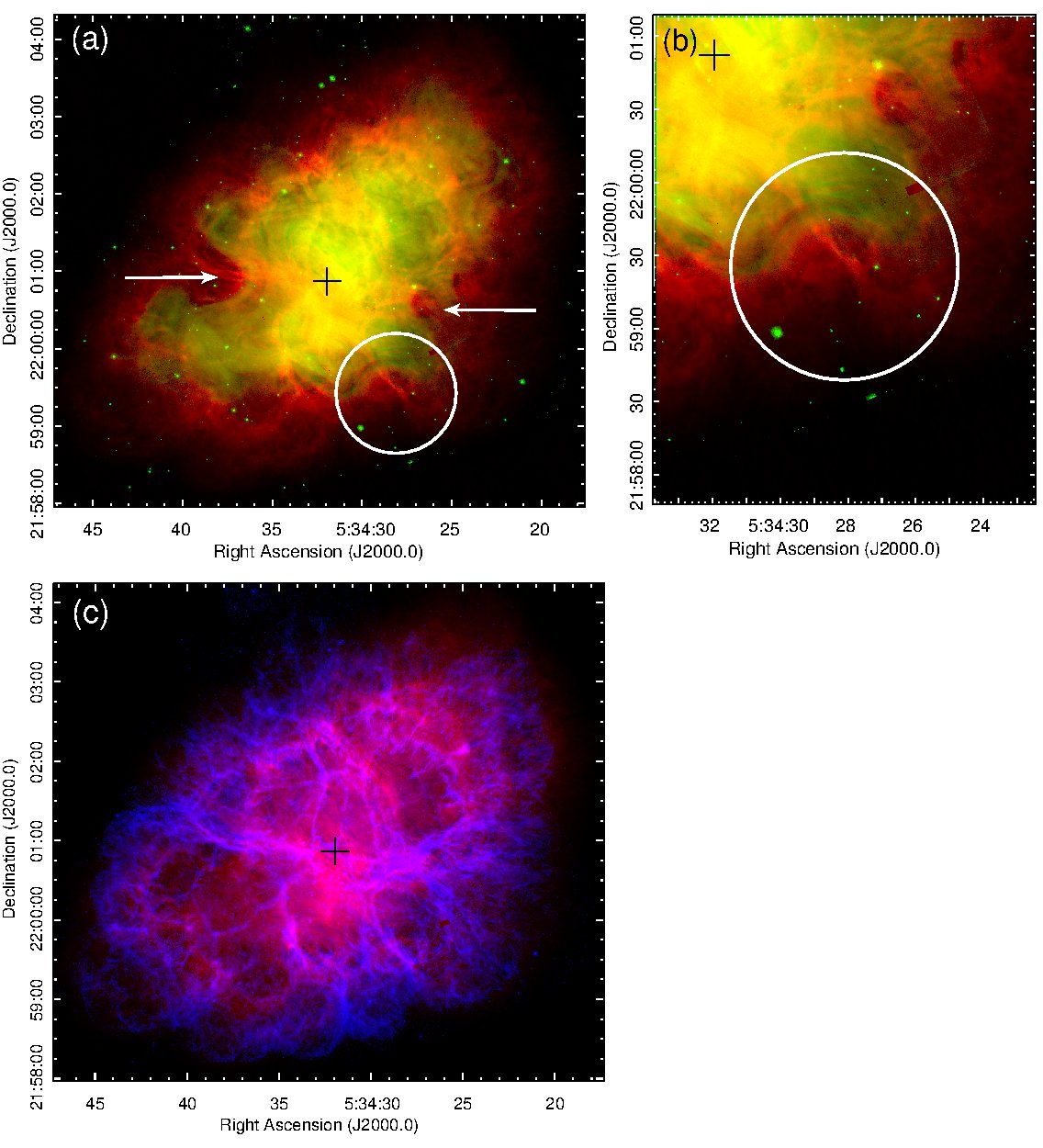}
\caption{{\bf (a)} Composition of radio emission at 3~GHz (in red, displayed with a linear scale between 0.05 and 5.3~mJy~beam$^{-1}$) with the optical continuum (in green) as observed in the filter F547M.  The white arrows point out the eastern and western ``bays'' discussed in the text. The white circle marks the ``arcade'' region. {\bf (b)} Zoom around the ``arcade-like'' structure.~{\bf (c)} Comparison of the radio emission at 3~GHz (in red) with the [OIII] $\lambda$5007 image (in blue). This combination highlights the fact that the radio synchrotron nebula is bounded by the region with high-ionization optical emission lines. The plus sign shows the location of the pulsar.  The radio image used is the spatially high-pass filtered 
one to emphasize the fine structure 
(see Section~\ref{observations}).
\label{radio-visible}}
\end{figure}

In Figure~\ref{radio-visible}(a) we present the comparison of distribution of the radio emission (in red) with the synchrotron-emitting optical continuum (in green) over the entire nebula. The optical synchrotron emission looks smoothly distributed and it appears to be completely confined within the radio nebula, as expected from the shorter lifetimes of the particles emitting in the visible range. Analogous to the IR synchrotron emission, the presence of the optical bays surrounded by radio emission with identical morphology can be noticed in Figure~\ref{radio-visible}(a), again with a small shift owing to the time lapse between both observations.
Another noticeable peculiarity is the distribution of the optical emission  related to the ``arcade'' (shown in Figure~\ref{radio-details}(b) and Figure~\ref{comparison-radio-58-8}). The region is zoomed in Figure~\ref{radio-visible}(b) to display the structure in detail. As already noticed in the previous section, it is remarkable how the synchrotron emission, in the optical range in this case,  exactly reproduces the radio arcade-like morphology around the straight filament labeled E. 

Figure~\ref{radio-visible}(c) displays in two colors a composition of the radio emission (in red) and the high-ionization [OIII] filamentary structure (in blue).
The high-ionization emission lines trace
the interface between the synchrotron nebula and the ejecta and are
part of what  \citet{hester1996} named the ``skin,'' consisting of swept-up thermal
ejecta that are being accelerated by the pressure of the synchrotron plasma
\citep{hester1996, sankrit1998, hester2008}.
As noticed before, the optical emission lines define the outer edge of the Nebula around most
of its periphery except for the northwestern portion \citep{loll2013}. From the present comparison
carried out with a very sensitive radio image, we confirm that the optical radiative shock at the boundary defines the outermost
border of the  Crab Nebula \citep[][and references therein]{hester2008}.

\begin{figure}
\plotone{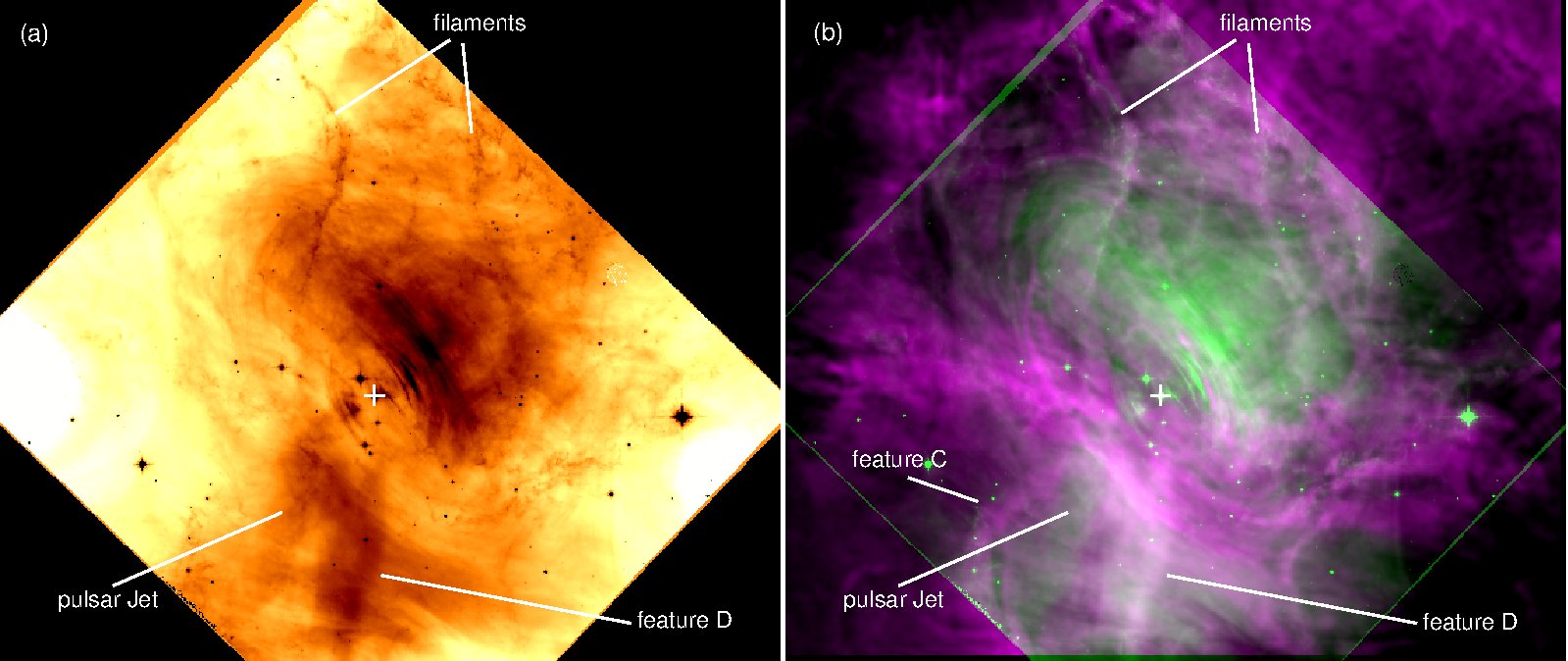}
\caption{{\bf (a)} {\it HST} WFC3/NIR image at 1.5$\mu$m of the central region of the Crab Nebula.  {\bf (b)} Comparison of the radio emission at 3~GHz (in magenta) with {\it HST} near-IR emission. The plus sign shows the location of the pulsar. Labeled features are discussed in the text.  The radio image used is the spatially high-pass filtered one to emphasize the fine structure (see Section~\ref{observations}).
\label{radio-HST}}
\end{figure}

For the region around the pulsar, where  small-scale features are highly variable, we performed the multiwavelength comparison using the {\it HST} data acquired on the same day as the radio one (about 11 hr later).  In Figure~\ref{radio-HST}(a) we present the {\it HST} WFC3/NIR image at 1.5$\mu$m. The well-known narrow wisps are more prominent to the northwest of the pulsar,   tracing concentric ellipses around the pulsar. Also,  a few narrow filaments of different extensions (indicated in the figure) can be seen running from the central plane to the north. As it will be shown below, these filaments have clear radio counterparts.  In the southern half the most conspicuous features are two broad structures, the one to the east being the optical counterpart of the pulsar jet, and the one to the west the counterpart of the feature labeled D in the radio image (Figure\ref{equatorial-region}(a)). It can be noticed that none of them have their starting point at the pulsar. The pulsar jet is visible approximately from the inner elliptical ring, while feature D seems to originate at a site about 12\s~ south (see also Figure~\ref{simultaneo}).

In Figure~\ref{radio-HST}(b) we present the superposition of the radio nebula (in magenta) with the same NIR image (in green). The regions where both emissions overlap are white. Although, as already noted, there is no a close correspondence of features,  some agreement can be noticed in the region adjacent to the pulsar. In the first place, the distribution of the radio wisps follows quite accurately the elliptical shape as traced by the NIR emission, although they do not overlap. In the inner ring, to the northwest of the pulsar, the radio wisps appear to shine in between NIR wisps. The denoted  ``filaments'' in Figure~\ref{radio-HST}(a) and (b), are conspicuous in both spectral ranges. To the south, we can notice that the radio-jet-like feature labeled C (Figure~\ref{equatorial-region}(a)) does not have an NIR counterpart (it is only seen in magenta color), while the pulsar jet (which is also bright in X-rays) does not have a radio counterpart (only seen in green color). Feature  D appears emitting both in radio and NIR. These last three structures will be considered again in the context of the X-ray emission analysis.

\section {Radio and ultraviolet}
\label{ultraviolet}

\begin{figure}
\plotone{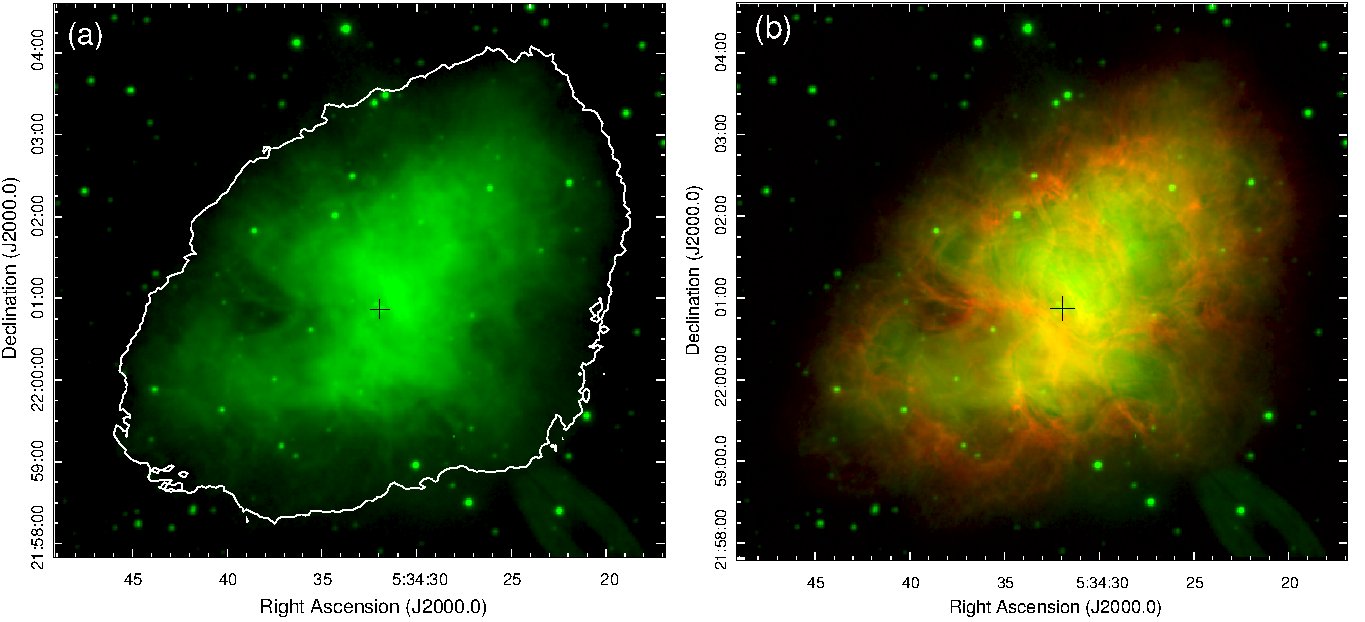}
\caption{{\bf (a)} Crab Nebula as observed in UV (around 291 nm) by the Optical Monitor on board  XMM-{\it Newton}. To illustrate the extension of the UV emission, a radio contour traced at 0.4~mJy~beam$^{-1}$ is plotted. {\bf (b)} Comparison of 3~GHz radio nebula (in red) with the UV emission (in green). The plus sign shows the location of the pulsar. The radio image used is the spatially high-pass filtered one  to emphasize the fine structure (see Section~\ref{observations}).
\label{UV}}
\end{figure}

Not many studies in the UV of the Crab Nebula have been published to date
\citep[e.g.][]{davidson1982, blair1992, hennessy1992, sollerman2000}, and they were mainly dedicated to spectroscopy of the brightest filamentary structure and/or selected regions. In a UV band comparable to the present data, it can be mentioned that  \citet{hennessy1992}  obtained images of the entire nebula in UV  at 150, 160, and 250 nm by  using  the Ultraviolet Imaging Telescope in the Astro-1 Shuttle, with  FWHM of $\sim$ 5\s, 
$\sim$ 3\s.6 and $\sim$ 4\s, respectively. Figure~\ref{UV}(a) shows  a new image of the Crab Nebula in UV wavelengths around 291 nm, obtained with an FWHM of 1\farcs7.

From Figure~\ref{UV} we see that the equatorial features are brighter northward of the pulsar, as in  the rest of investigated wavelengths. Only hints of the southern jet can be traced, somewhat confused with more diffuse emission\footnote{If the UV jet moves around as the X-ray jets does, then the image may be blurred because of averaging over a large time interval}. The eastern and western bays are very conspicuous features in the UV. With the appropriate contrast it can be noticed that the UV radiation covers an area  almost as extensive as the radio nebula.  In Figure~\ref{UV}(b) we present a two-color image combining radio emission (in red) with the  UV  (in green). In particular, radio and UV emissions overlap in features A, B, D, and E, in the equatorial region. Based on the extension and great morphological similitude with the continuum optical and infrared  emission, we suggest that the  UV emission shown in Figure~\ref{UV} is probably of synchrotron origin.

\section {Radio and X-rays}
\label{xrays}
\begin{figure}
\plotone{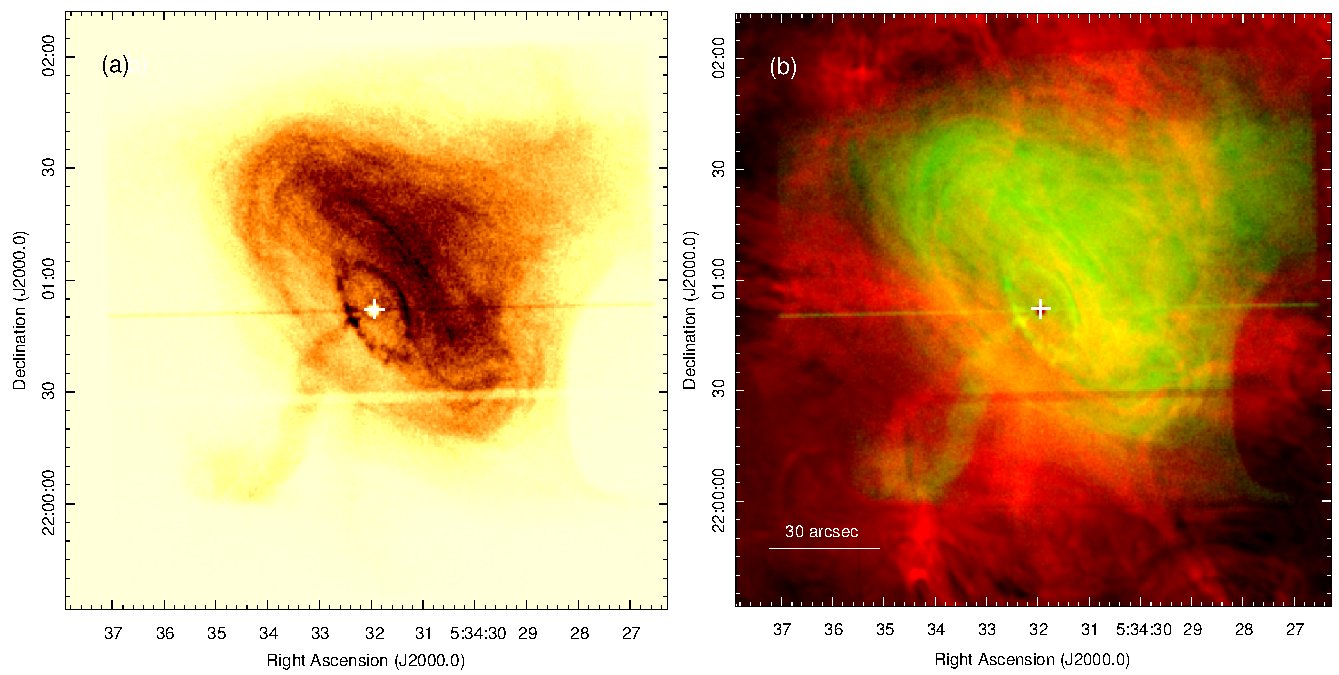}
\caption{{\bf (a)} \it Chandra \rm X-ray image in the 0.5-8~keV band as obtained in the 2012 campaign.
{\bf (b)} The same X-ray image (in green) overlapping the 3~GHz radio emission (in red).
The plus sign shows the location of the pulsar in both images. The radio image used is the spatially
high-pass filtered  one described in Section~\ref{observations} to emphasize the fine structure.
\label{radio-xrays}}
\end{figure}

Figure~\ref{radio-xrays}(a) shows the X-ray image of the central region of the Crab Nebula as obtained during
the 2012 \it Chandra \rm campaign. It shows the knotty bright inner ring
\citep[the ``inner ring'' described by][]{weisskopf2012}, the elliptical torus with bright wisps, and the
jet extending to the southeast. The southeastern jet extends about 17$^{\prime\prime}$
in a more or less straight way along the spin axis direction from the inner ring up to the position
$\sim$ 5\hh 34\mm 33\ss, +22\d 00\m 40\s, where it bends to the south extending for $\sim$ 31$^{\prime\prime}$,
and ends in a curl, near 5\hh 34.5\mm 33\ss, +22\d 00\m 10\s. The jet appears well collimated only
in the first section, and then it gets progressively fainter and broader with distance. Multiyear \it Chandra \rm observations show that the X-ray jet changes its shape \citep{weisskopf2012}, which could
be explained by either precession of the curved jet or by the motion of kinks along it
\citep[][and references therein]{kargaltsev2015}.

To compare in detail the X-ray radiation with the radio emission, we present in 
Figure~\ref{radio-xrays}(b)
the superposition of the present \it Chandra  \rm  X-ray image with the VLA 3~GHz image. The most striking
feature of this figure is the presence of two separate jet-like features, one emitting in X-rays (in green)
and the other in radio\footnote{The presence of this structure was previously communicated by
\citet{kargaltsev2015} based on the same VLA image presented here.} (in red), corresponding to the feature labeled C
(Figure~\ref{equatorial-region}(a)). From the comparison of the new 3~GHz radio image with that at 5~GHz obtained in 2001 and
reported by \citet{bietenholz2004}, we conclude that the ``radio-jet" did not change along the 11 yr time
span, neither in shape nor in position.

The X-ray jet and the ``radio-jet'' do not start at the same distance from the pulsar. While the
X-ray jet appears to have its starting point at the innermost ring around the pulsar, the ``radio-jet''
starts at a larger distance from the pulsar, near the radio feature that we identified as ``A'' in
Figure~\ref{equatorial-region}(a). Sky projections of the radio and X-ray jet-like features cross at the point
where the X-ray emission makes the curl.
After that point the radio emission broadens and bends to the south, extending for about 43$^{\prime\prime}$
where it ends in a small, bright knot at 5\hh 34\mm 34\ss, +21\d 59\m 40\s. The two jet-like features
could be unrelated structures whose projections overlap in the plane of the sky. Alternatively, both
the X-ray and radio jets might be due to synchrotron emission of relativistic particles supplied by
the pulsar. In this case we should understand why the radio-emitting particles follow a different path
than the X-ray-emitting particles.

\section{Simultaneous View of the Ventral PWN in Three Spectral Bands}

\begin{figure}
\centering
\plotone{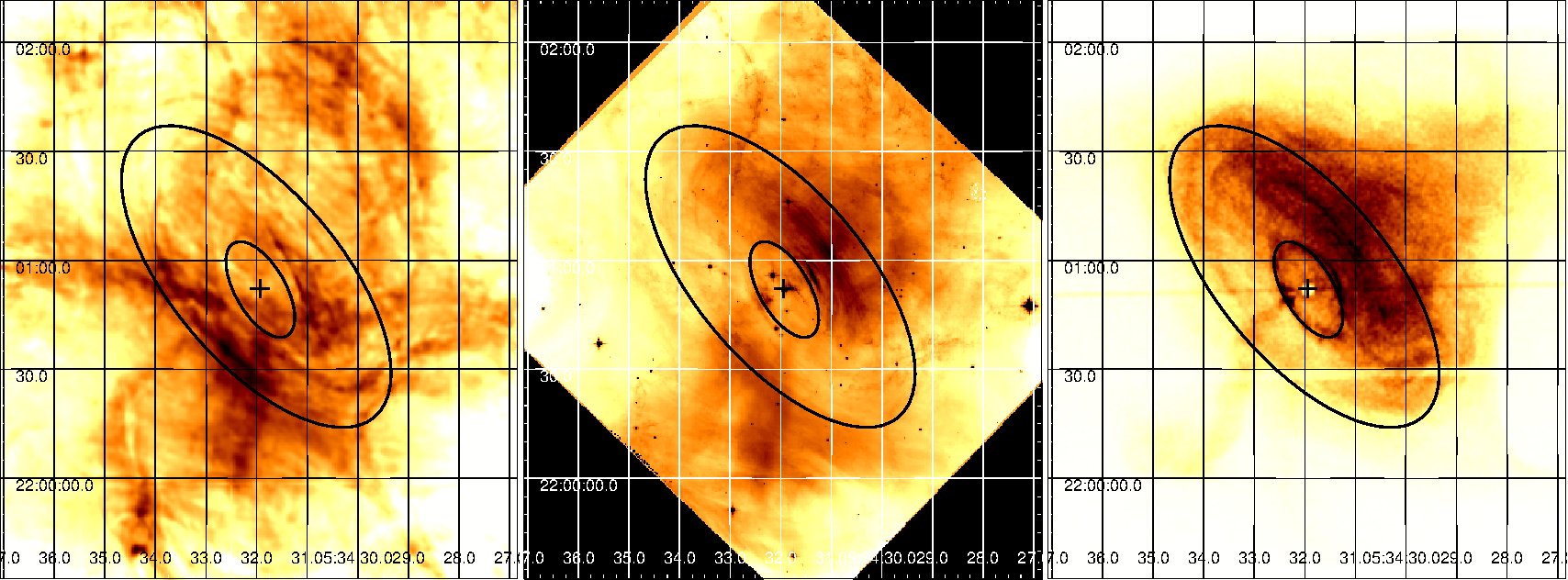}
\caption{From left to right: radio, NIR, and X-ray contemporaneous images of the center of the Crab Nebula. An inverted color scheme is used to improve the display of features (darker is brighter). The plus sign shows the location of the pulsar. The radio image used is the spatially high-pass filtered one (see Section~\ref{observations}) to emphasize the fine structure. The overplotted ellipses are discussed in the text.
\label{simultaneo}}
\end{figure}

For a quick view of similarities  and differences in the appearance of the central PWN,
in Figure~\ref{simultaneo} we show the three 2012 November 26 images (from left to right: the
central region of the Crab Nebula is shown in radio, near-IR, and X-rays).
To help the comparison we have overplotted the coordinates grid, an outer ellipse that
approximately depicts the toroidal structure around the pulsar, and an internal ellipse
coincident with the inner X-ray knotty ring.
This figure confirms that even the bright wisps located northwest of the pulsar (the only
features around the pulsar that have the same shape in the different spectral ranges) do not
match in any of the images shown, the greatest differences being found with the radio emission.
Also, from Figure~\ref{simultaneo}, it is readily apparent that while in the radio band the brightest
features are preferentially located southeast of the pulsar, at higher energies the northwestern
side is more prominent.

To explain the lack of spatial coincidence between radio features and the radiation in the optical
and X-rays ranges in the pulsar torus, \citet{komissarov2013} proposed that the radio wisps could
just be some kind of ripples driven by the unsteady outflow from the termination shock through the PWN.
In this case, there should be delayed changes in the radio wisp positions, which would correlate with
the changes in the X-ray ring shape/brightness. To look for these correlations one needs to carry out
radio follow-up of X-ray monitoring observations on longer timescales.
An alternative explanation, also suggested by \citet{komissarov2013}, is that the radio particles
come from different parts of the termination shock. The existence of two synchrotron components of
different origin was earlier proposed by \citet{bandiera2002} to explain spectral peculiarities between
the radio emission at 20~cm and at 1.3~mm. Also, \citet{schweizer2013} proposed that the X-ray and the
optical emission must be produced by different populations of particles. Further investigations are
required to discern the origin and distribution of the synchrotron radiation in the different spectral ranges.

\section{Summary}
\label{conclusions}

We have presented a completely new set of images of the Crab Nebula, from radio to X-rays, including a VLA radio image at 3~GHz with subarcsecond angular resolution (entire nebula), the first ALMA image at 100~GHz (central region), an {\it HST} WFPC3 F160W NIR image (central region), an {\it XMM}-OM UV image around 291~nm (entire nebula), and a {\it Chandra} ACIS X-ray image (central region). Based on these images and previous {\it Spitzer} IR and {\it HST} optical images of the Crab Nebula, we carried out a detailed analysis of the morphological properties of the nebula across the electromagnetic spectrum, with particular emphasis on the central region where we compared almost simultaneous images.

Our main findings are as follows:

1. The radio emission at 3~GHz shows the well-known undulated border.  Particularly, toward the east, southeast, and south, well-defined arches composed of a series of narrow filaments dominate the edge appearance. Along the eastern boundary these filaments form large curved features with footpoint brightenings that we interpret as produced by particle acceleration and magnetic enhancement, in an analogy with solar phenomena that produce a similar topology. Also, in the southwest part of the nebula a set of closely packed arches can be observed wrapping around a long filament that comes from the equatorial region, with a morphology resembling the solar ``arcades'' formed by closely occurring loops of magnetic lines of force. We suggest that the loop-like structures originate in plasma confined to magnetic field lines, and the magnetic fields are restructured by plasma kink instabilities, which change the magnetic structure from the central toroidal geometry with nested loops, into a chaotic structure with loops throughout the nebula, as proposed by \citet{begelman1998}.

2. A preliminary look at the central region of the nebula as observed for the first time at 100~GHz with angular resolution better than 2\s, suggests that the radio synchrotron emission has the same spatial distribution up to millimeter wavelengths, a frequency where the emitting electrons are over five times more energetic than those emitting at 3~GHz. More observations of the entire nebula are required to confirm the results.

3. The continuum synchrotron emission of the Crab Nebula as mapped with {\it Spitzer} in IR, in optical with the {\it HST}, and in the new {\it XMM}-OM UV image, shows two very conspicuous ``bays,'' one to the east and one to the west. They are peripheral indentations of the Crab Nebula probably produced by a pre-supernova disklike magnetized torus that blocks the penetration of relativistic particles. From the present comparison with the radio image, it is possible to confirm that the bays are also a morphological property of the radio emission (the east bay being better defined). The richness of the borderline features masks these indentations in the radio regime.  Also, the present multiwavelength comparison revealed that the ``arcade-like'' feature observed in radio is clearly detected in IR (at 4.5  $\mu$m that traces the synchrotron component) and in the optical continuum.

4. The comparison of the high-dynamic range 3~GHz image with the {\it HST} mosaic obtained in the [OIII] line \citep [from][]{loll2013}, confirms that the swept-up thermal ejecta accelerated by the pressure of the PWN plasma traces the outer edge of the Nebula around most of its periphery, with the only exception for the northwestern portion.

5. The new {\it HST} NIR image of the central region shows the well-known elliptical torus around the pulsar, composed of a series of concentric narrow features of variable intensity and width. The emission is in general more prominent northwest of the pulsar.

6. The comparison of the radio and the X-ray emission distributions in the central region suggests the existence of a double-jet system from the pulsar, one detected in X-rays and the other in radio. None of them starts at the pulsar itself but in its environs. The X-ray jet begins at the bright inner ring, while what we call the ``radio-jet" appears to start in a radio feature located about 5\farcs5 ($\sim 1.6 \times 10^{17}$~cm at a distance of 2~kpc) southeast from the pulsar. Contrary to the X-ray jet whose shape changes with time, from the comparison of the new 3~GHz image with previous radio images we conclude that the ``radio-jet'' looks constant in shape and position along the years.

7. The comparison of simultaneous multiwavelength emission in the vicinity of the pulsar suggests the existence of at least two different synchrotron components.

\acknowledgments
G.D. and G.C.  are members of the {\it Carrera del Investigador Cient\'\i fico} of CONICET, Argentina. We thank C. Mandrini for the useful discussions about solar phenomena.
This work was partially supported by grants awarded by CONICET (PIP 0736/11) and  ANPCYT (PICT 0571/11), Argentina. O.K. acknowledges partial support from the Chandra Award GO3-14084X issued by the Chandra X-ray Observatory Center, and Hubble Space Telescope award HST-GO-13043.009-A issued by the Space Telescope Science Institute.
This paper makes use of the following ALMA data:
   ADS/JAO.ALMA\#2012.1.01099.S. ALMA is a partnership of ESO (representing  its member states), NSF (USA) and NINS (Japan), together with NRC
   (Canada) and NSC and ASIAA (Taiwan), in cooperation with the Republic of Chile. The Joint ALMA Observatory is operated by ESO, AUI/NRAO and NAOJ. This research has made use of SAOImage DS9, developed by Smithsonian 
Astrophysical Observatory.

\nocite{*}
\bibliographystyle{aasjournal} 
\bibliography{bib_crab}

\end{document}